# Asynchronous Early Output Dual-Bit Full Adders Based on Homogeneous and Heterogeneous Delay-Insensitive Data Encoding

P. BALASUBRAMANIAN[1*], K. PRASAD[2]
[1] School of Electrical and Electronic Engineering
Nanyang Technological University
50 Nanyang Avenue
Singapore 639798
balasubramanian@ntu.edu.sg
[2] Department of Electrical and Electronic Engineering
Auckland University of Technology
Auckland 1142
New Zealand
krishnamachar.prasad@aut.ac.nz

*Abstract:* - This paper presents the designs of asynchronous early output dual-bit full adders without and with redundant logic (implicit) corresponding to homogeneous and heterogeneous delay-insensitive data encoding. For homogeneous delay-insensitive data encoding only dual-rail i.e. 1-of-2 code is used, and for heterogeneous delay-insensitive data encoding 1-of-2 and 1-of-4 codes are used. The 4-phase return-to-zero protocol is used for handshaking. To demonstrate the merits of the proposed dual-bit full adder designs, 32-bit ripple carry adders (RCAs) are constructed comprising dual-bit full adders. The proposed dual-bit full adders based 32-bit RCAs incorporating redundant logic feature reduced latency and area compared to their non-redundant counterparts with no accompanying power penalty. In comparison with the weakly indicating 32-bit RCA constructed using homogeneously encoded dual-bit full adders containing redundant logic, the early output 32-bit RCA comprising the proposed homogeneously encoded dual-bit full adders with redundant logic reports corresponding reductions in latency and area by 22.2% and 15.1% with no associated power penalty. On the other hand, the early output 32-bit RCA constructed using the proposed heterogeneously encoded dual-bit full adder which incorporates redundant logic reports respective decreases in latency and area than the weakly indicating 32-bit RCA that consists of heterogeneously encoded dual-bit full adders with redundant logic by 21.5% and 21.3% with nil power overhead. The simulation results obtained are based on a 32/28nm CMOS process technology.

## 1 Introduction

The full adder forms the fundamental component of arithmetic circuits used in various microprocessor, microcontroller and digital signal processor based applications. The full adder is basically used to add two binary inputs along with a carry input from a preceding stage and produces two binary outputs viz. sum and carry output (also called as carry overflow). The full adder can be realized in either synchronous [1] – [4] or asynchronous design style [5] – [15]. As an alternative to the conventional single-bit full adder (SBFA) the concept of a dual-bit full adder (DBFA) was proposed in [16] – [18] based on the synchronous and asynchronous design paradigms. The DBFA adds two augend and addend binary inputs along with any carry input and produces two sum outputs along with the carry overflow. It was shown in [16] – [18] that regardless of whether the circuit designs are synchronous or asynchronous, the DBFA when cascaded to form a ripple carry adder (RCA) would help to substantially reduce the latency (i.e. critical path delay) of a RCA constructed using SBFAs albeit at the expense of some area and power overheads. Nevertheless, the power-delay and/or energy-delay products tend to remain optimized. Moreover, it was pointed out that a hybrid design involving DBFAs

---

and SBFAs could be beneficial in terms of further optimizing power, delay and area although this may be a peephole optimization strategy.

In this work, we present the novel designs of two asynchronous early output DBFAs based on homogeneous and heterogeneous delay-insensitive data encoding without and with redundant logic. We show that the proposed designs report considerably less latency, area and power dissipation than the previously proposed asynchronous DBFAs when incorporated into a RCA architecture. This inference is based on simulations performed using a 32/28nm CMOS process. When comparing the latency, area and power metrics of SBFA based RCA counterparts with the design metrics of DBFA based RCAs for a 32-bit addition operation, we infer that the proposed asynchronous early output DBFAs based RCAs which incorporate redundant logic report the least latency amongst all. Nonetheless, the latencies of DBFAs based asynchronous RCAs can be further reduced through hybrid designs which involve both asynchronous DBFAs and SBFAs.

The remainder of this research paper is organized as follows. Some relevant background about robust asynchronous design based on delay-insensitive data codes, homogeneous and heterogeneous delay-insensitive data encoding, and the 4-phase return-to-zero handshake protocol is provided in Section 2. The proposed designs of the asynchronous early output DBFAs corresponding to homogeneous and heterogeneous delay-insensitive data encoding are presented in Section 3. Next, the simulation results of various 32-bit asynchronous RCAs utilizing diverse DBFAs are given in Section 4. Lastly, Section 5 draws the conclusions.

## 2 Asynchronous Design – Background

An asynchronous function block is the equivalent of the synchronous combinational logic [19]. When an asynchronous function block is constructed using delay-insensitive codes [20] and utilizes a 4-phase handshaking, it is generally robust provided it is free of gate and wire orphans [21] – [23]. Orphans are unacknowledged signal transitions which may occur on gate outputs (i.e. gate orphans) or wires (i.e. wire orphans). Wire orphans are usually eliminated by imposing the isochronicity assumption [24], which is the weakest compromise to delay-insensitivity. An isochronic fork implies that a signal transition on a wire junction (i.e. node) is concurrently transmitted on all the wire branches. However, gate orphans may become problematic and hence their possibility of occurrence should be neutralized to guarantee that an asynchronous design remains robust.

The dual-rail code (also called 1-of-2 code) is the simplest member of the family of delay-insensitive $m$-of-$n$ data codes [20]. Among the family of $m$-of-$n$ codes, 1-of-$n$ codes represent a subset and are called one-hot codes. In a 1-of-$n$ code, only 1 out of $n$ wires is asserted high (i.e. binary 1) to represent a binary data. In fact, the 1-of-$n$ coding scheme is said to be unordered [25] since none of the code words forms a subset of another code word. Also, the 1-of-$n$ coding scheme is said to be complete [26] if all the $n$ unique code words, as per definition, are utilized to encode the specified binary data. Table 1 shows an example binary data representation according to the 1-of-2 and 1-of-4 data encoding schemes.

Table 1. Example 2-bit binary data representation in 1-of-2 and 1-of-4 data encoding schemes

| Binary data | | 1-of-2 encoded data | | 1-of-4 encoded data | | | |
|---|---|---|---|---|---|---|---|
| X | Y | (X1,X0) | (Y1,Y0) | E0 | E1 | E2 | E3 |
| 0 | 0 | (0,1) | (0,1) | 1 | 0 | 0 | 0 |
| 0 | 1 | (0,1) | (1,0) | 0 | 1 | 0 | 0 |
| 1 | 0 | (1,0) | (0,1) | 0 | 0 | 1 | 0 |
| 1 | 1 | (1,0) | (1,0) | 0 | 0 | 0 | 1 |

As per the 1-of-2 code, a single-rail binary input, say D, is encoded using two wires, say D1 and D0, where the data D = 1 is represented by D1 = 1 and D0 = 0, and the data D = 0 is represented by D1 = 0 and D0 = 1. Note that both D1 and D0 cannot assume 1 simultaneously as it is illegal and invalid because the coding scheme will no more be unordered. However, both D1 and D0 can assume 0 simultaneously and is referred to as the spacer. Hence as per the 1-of-2 code a valid data is specified by either D1 or D0 assuming binary 0 and the other assuming binary 1, and the condition of both D1 and D0 assuming binary 0 is labelled as the spacer or null (i.e. empty data). On the other hand, the 1-of-4 code is used to represent two bits of binary information at a time. Referring to Table 1, it can be seen that the two binary inputs specified by X and Y are encoded into E0, E1, E2 and E3 as per the 1-of-4 code for an illustration.

When just one delay-insensitive code (say, 1-of-2 code) is alone used to encode the given binary data, it is called homogeneous data encoding, and when more than one delay-insensitive code (for example, 1-of-2 and 1-of-4 codes) is used to encode the given binary data, it is called heterogeneous data encoding.

A typical asynchronous system stage that employs delay-insensitive codes for data encoding and data processing and the 4-phase return-to-zero handshake protocol for data communication is shown in Fig 1. As the name suggests, the 4-phase return-to-zero handshake protocol consists of 4 phases. This will be explained with reference to Fig 1 based on the assumption that the 1-of-2 code is used for data representation. Nevertheless, the explanation would be applicable for data representation using any delay-insensitive 1-of-$n$ code.

In the first phase, the dual-rail data bus shown in Fig 1 is in the spacer state and ACKIN is high i.e. binary 1. The transmitter now transmits a code word i.e. valid data and this results in upgoing signal transitions on any one of the corresponding dual rails of the entire dual-rail data bus. In the second phase, the receiver receives the code word sent, and it drives ACKOUT high. In the next phase viz. third phase, the transmitter waits for ACKIN to go low i.e. binary 0 and then resets the entire dual-rail data bus to spacer state. Subsequently, in the fourth phase, after an unbounded time duration, which is finite and positive though, the receiver drives ACKOUT low i.e. ACKIN becomes high. One data transaction is now said to be completed and the asynchronous system stage is ready to commence the next data transaction.

The completion detector [19] shown in Fig 1 ensures the complete arrival of all the primary inputs into an asynchronous system stage whether they are valid data or spacer. It consists of an array of 2-input OR gates in the first logic level with each 2-input OR gate used to combine the respective dual-rails of an encoded primary input. The outputs of all the 2-input OR gates are synchronized using a C-element[†] tree, whose granularity depends on the composition of the digital cell library used for physical implementation.

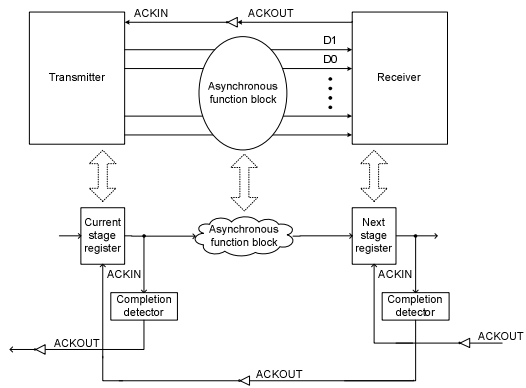

Fig 1 A robust asynchronous system stage operation correlated with the transmitter-receiver analogy

Asynchronous function blocks are generally classified as strongly indicating, weakly indicating and early output types. Indication basically means acknowledging the arrival of the inputs to a circuit or system through corresponding monotonic transitions on the intermediate and primary outputs, where the transitions should be either monotonically increasing or decreasing uniformly throughout the entire circuit or system [27]. The generalized input-output timing characteristics of strong-indication, weak-indication and early output type asynchronous function blocks are captured through Fig 2.

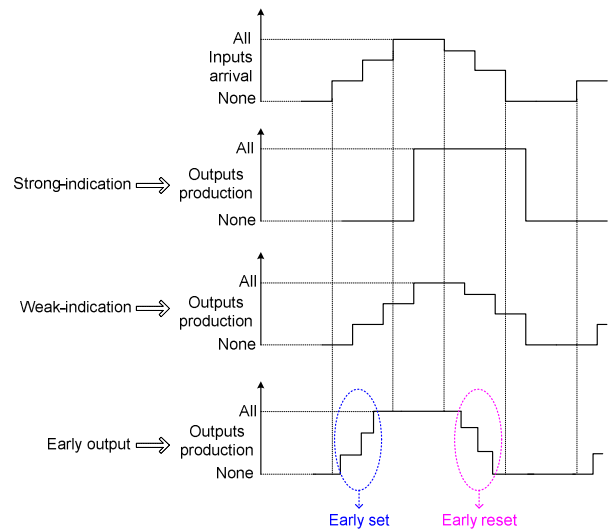

Fig 2 Depicting inputs-outputs timing correlation of strong-indication, weak-indication and early output asynchronous function blocks

A strong-indication function block [5] [28] starts data processing only after receiving all the primary inputs, and the requisite outputs are then produced. A weak-indication function block [5] [29] is able to commence data processing after receiving just a subset of the primary inputs and can also produce some primary outputs. However, only after receiving the last primary input, the last corresponding primary output is produced by the weak-indication function block. With respect to indication, the mechanism may be either local or global [30]: local, if the asynchronous function block is internally indicating, and global, if the asynchronous system stage provides indication externally. It was shown in [30] that local indication is preferable over global indication for robust asynchronous circuit designs.

An early output function block [31] [32] is in fact the most relaxed compared to strong-indication or weak-indication function blocks as it can commence data processing after receiving just a subset of the primary inputs and subsequently can produce all the primary outputs without waiting for the arrival of all the primary inputs. In this context, the early output function block could exhibit early set or early reset behavior as highlighted in Fig 2. Early set implies that upon receiving a subset of the valid data (primary) inputs, the early output function block produces all the valid data (primary) outputs. The early set property is highlighted through the blue oval in Fig 2. On the other hand, early reset implies that upon receiving a subset of spacer data (primary) inputs, the early output function block processes them and drives all the primary outputs to the spacer state. The early reset property is highlighted through the pink oval in Fig 2.

---

[†] The C-element is basically a rendezvous element. If all its inputs are binary 1 or 0, it outputs binary 1 or 0 respectively. However, if its inputs are different, the C-element retains its existing output.

## 3 Proposed Asynchronous DBFAs

Novel asynchronous DBFAs based on homogeneous and heterogeneous delay-insensitive data encoding were designed without and with redundant logic (which is implicit), and they are described next. For homogeneous data encoding, 1-of-2 code is used and for heterogeneous data encoding, 1-of-2 and 1-of-4 codes are used.

### 3.1 Homogeneously Encoded Early Output Asynchronous DBFAs

In the case of the homogeneously encoded DBFAs, only the 1-of-2 code is used for encoding the augend and addend inputs, the carry input, the carry output, and the sum outputs. Let (A11, A10) and (A01, A00)

$$
\begin{aligned}
\text{COUT1} = {} & \text{A10A00B11B01CIN1} + \text{A11A00B10B01CIN1} + \text{A10A01B11B00CIN1} \\
& + \text{A11A01B10B00CIN1} + \text{A10A01B11B01} + \text{A11A01B10B01} + \text{A11B11} \quad (1)
\end{aligned}
$$

$$
\begin{aligned}
\text{COUT0} = {} & \text{A11A01B10B00CIN0} + \text{A10A01B11B00CIN0} + \text{A11A00B10B01CIN0} \\
& + \text{A10A00B11B01CIN0} + \text{A11A00B10B00} + \text{A10A00B11B00} + \text{A10B10} \quad (2)
\end{aligned}
$$

$$
\begin{aligned}
\text{SUM11} = {} & \text{A11A01B10B00CIN0} + \text{A10A01B11B00CIN0} + \text{A11A00B10B01CIN0} \\
& + \text{A10A00B11B01CIN0} + \text{A11A00B11B01CIN1} + \text{A11A01B11B00CIN1} \\
& + \text{A10A00B10B01CIN1} + \text{A10A01B10B00CIN1} + \text{A10A01B10B01} \\
& + \text{A11A00B10B00} + \text{A10A00B11B00} + \text{A11A01B11B01} \quad (3)
\end{aligned}
$$

$$
\begin{aligned}
\text{SUM10} = {} & \text{A11A01B10B00CIN1} + \text{A10A01B11B00CIN1} + \text{A11A00B10B01CIN1} \\
& + \text{A10A00B11B01CIN1} + \text{A10A01B10B00CIN0} + \text{A10A00B10B01CIN0} \\
& + \text{A11A01B11B00CIN0} + \text{A11A00B11B01CIN0} + \text{A11A00B11B00} \\
& + \text{A11A01B10B01} + \text{A10A01B11B01} + \text{A10A00B10B00} \quad (4)
\end{aligned}
$$

$$
\text{SUM01} = \text{A01B00CIN0} + \text{A00B01CIN0} + \text{A00B00CIN1} + \text{A01B01CIN1} \quad (5)
$$

$$
\text{SUM00} = \text{A01B01CIN0} + \text{A01B00CIN1} + \text{A00B01CIN1} + \text{A00B00CIN0} \quad (6)
$$

---

$$
\begin{aligned}
\text{COUT1} = {} & \text{A0B3CIN1} + \text{A1B2CIN1} + \text{A2B1CIN1} + \text{A3B0CIN1} + \text{A1B3} + \text{A2B2} \\
& + \text{A3B1} + \text{A2B3} + \text{A3B2} + \text{A3B3} \quad (7)
\end{aligned}
$$

$$
\begin{aligned}
\text{COUT0} = {} & \text{A0B3CIN0} + \text{A1B2CIN0} + \text{A2B1CIN0} + \text{A3B0CIN0} + \text{A0B0} + \text{A0B1} \\
& + \text{A0B2} + \text{A1B0} + \text{A1B1} + \text{A2B0} \quad (8)
\end{aligned}
$$

$$
\begin{aligned}
\text{SUM3} = {} & \text{A0B3CIN0} + \text{A1B2CIN0} + \text{A2B1CIN0} + \text{A3B0CIN0} + \text{A0B2CIN1} \\
& + \text{A1B1CIN1} + \text{A2B0CIN1} + \text{A3B3CIN1} \quad (9)
\end{aligned}
$$

$$
\begin{aligned}
\text{SUM2} = {} & \text{A0B2CIN0} + \text{A1B1CIN0} + \text{A2B0CIN0} + \text{A3B3CIN0} + \text{A0B1CIN1} \\
& + \text{A1B0CIN1} + \text{A2B3CIN1} + \text{A3B2CIN1} \quad (10)
\end{aligned}
$$

$$
\begin{aligned}
\text{SUM1} = {} & \text{A0B1CIN0} + \text{A1B0CIN0} + \text{A2B3CIN0} + \text{A3B2CIN0} + \text{A0B0CIN1} \\
& + \text{A1B3CIN1} + \text{A2B2CIN1} + \text{A3B1CIN1} \quad (11)
\end{aligned}
$$

$$
\begin{aligned}
\text{SUM0} = {} & \text{A0B0CIN0} + \text{A1B3CIN0} + \text{A2B2CIN0} + \text{A3B1CIN0} + \text{A0B3CIN1} \\
& + \text{A1B2CIN1} + \text{A2B1CIN1} + \text{A3B0CIN1} \quad (12)
\end{aligned}
$$

represent the dual-rail augend inputs, and let (B11, B10) and (B01, B00) represent the dual-rail addend inputs. Also, let (CIN1, CIN0) represent the dual-rail carry input. The most significant and least significant dual-rail sum outputs are specified by (SUM11, SUM10) and (SUM01, SUM00) respectively. (COUT1, COUT0) represents the dual-rail carry output. The logic equations corresponding to the homogeneously encoded DBFA are given by (1) to (6). It may be noticed that all the DBFA outputs are expressed in the disjoint sum-of-products form [33]. In a disjoint sum-of-products form, the logical conjunction of any two products results in null [34] since the product terms are mutually orthogonal [35].

Fig 3 shows the synthesized early output asynchronous DBFA based on homogeneous data encoding, which is technology mapped to the 32/28nm CMOS cell library [36]. Fig 3 contains a mix of discrete gates, complex gates and custom-designed 2-input C-elements, which are symbolized through the circle with the marking 'C' on them. Since input-incomplete gates [23] are used in the proposed homogeneously encoded DBFA designs to process the primary data inputs in the first logic level, they correspond to early output i.e. early reset type.

If the two complex gates viz. AO21 gates shown within the red and blue rectangles in dotted lines in Fig 3 are removed, and if the two 2-input OR gates depicted in red and blue in dotted lines in Fig 3 are retained to synthesize COUT1 and COUT0 respectively, then the homogeneously encoded asynchronous DBFA portrayed by Fig 3 does not have logic redundancy [37], especially with respect to the carry output logic. Alternatively, if the two 2-input OR gates shown in dotted lines in red and blue are removed, and if the two complex gates shown within the red and blue rectangles in dotted lines in Fig 3 are retained, then the homogeneously encoded asynchronous DBFA is said to contain redundant logic [37]. However, logic redundancy is implicit in the design.

For the asynchronous DBFA shown in Fig 3 when positioned in an intermediate position in a RCA architecture, the elements present in the critical path of the non-redundant design would be a 2-input C-element and a 2-input OR gate. On the contrary, the element found in the critical path of the redundant design would be just the AO21 gate. Hence, it becomes evident that the latency of the RCA embedding the proposed homogeneously encoded DBFA with redundant logic would be less than the latency of the RCA containing the homogeneously encoded DBFA with no redundant logic. But logic redundancy may cause a slight increase in area in the case of the former compared to the latter.

## 3.2 Heterogeneously Encoded Early Output Asynchronous DBFAs

In the case of the heterogeneously encoded DBFAs, the 1-of-2 code is used to encode the carry input and the carry output, while the 1-of-4 code is used to encode the augend and addend inputs, and the sum outputs based on Table 1. The 1-of-4 encoded augend and addend inputs are denoted by A0, A1, A2, A3 and B0, B1, B2, B3 respectively. The 1-of-4 encoded sum outputs are denoted by SUM0, SUM1, SUM2 and SUM3. As mentioned earlier, (CIN0, CIN1) and (COUT0, COUT1) represent the dual-rail carry input and carry output respectively. The logical equations corresponding to the heterogeneously encoded asynchronous DBFA are specified by (7) to (12). Again, (7) and (12) are expressed in disjoint sum-of-products form, whose respective product terms are all mutually orthogonal. Fig 4 shows the proposed asynchronous early output DBFA corresponding to heterogeneous data encoding, which is synthesized using discrete, complex and custom-designed 2-input C-gates which are eventually technology mapped to the 32/28nm cell library [36].

In Fig 4, if the complex gates viz. AO21 gates shown within the pink and green rectangles in dotted lines are removed and if the two 2-input OR gates highlighted in pink and green in dotted lines are retained to produce COUT1 and COUT0 respectively then the asynchronous DBFA is said to have no redundant logic, especially with respect to the carry output logic. Alternatively, if the two 2-input OR gates highlighted in pink and green in dotted lines in Fig 4 are removed, and if the two AO21 gates shown within the pink and green rectangles in dotted lines are retained the asynchronous DBFA shown in Fig 4 is said to contain redundant logic. The critical data path of the heterogeneously encoded DBFA which has no redundant logic when present in an intermediate position in a RCA architecture consists of a 2-input C-element and a 2-input OR gate, while the critical data path of the heterogeneously encoded DBFA with redundant logic when present in a similar position in the RCA architecture comprises just a single AO21 gate. Hence the latency would be less in the case of the RCA constructed by cascading the proposed heterogeneously encoded DBFA with redundant logic than the latency of the RCA constructed by cascading the heterogeneously encoded DBFA with no redundant logic although the former may occupy slightly more area compared to the latter due to extra logic. Note that the proposed heterogeneously encoded asynchronous DBFAs use input-incomplete gates to process the primary data inputs in the first logic level in Fig 4 and hence they would exhibit early output i.e. early reset behavior.

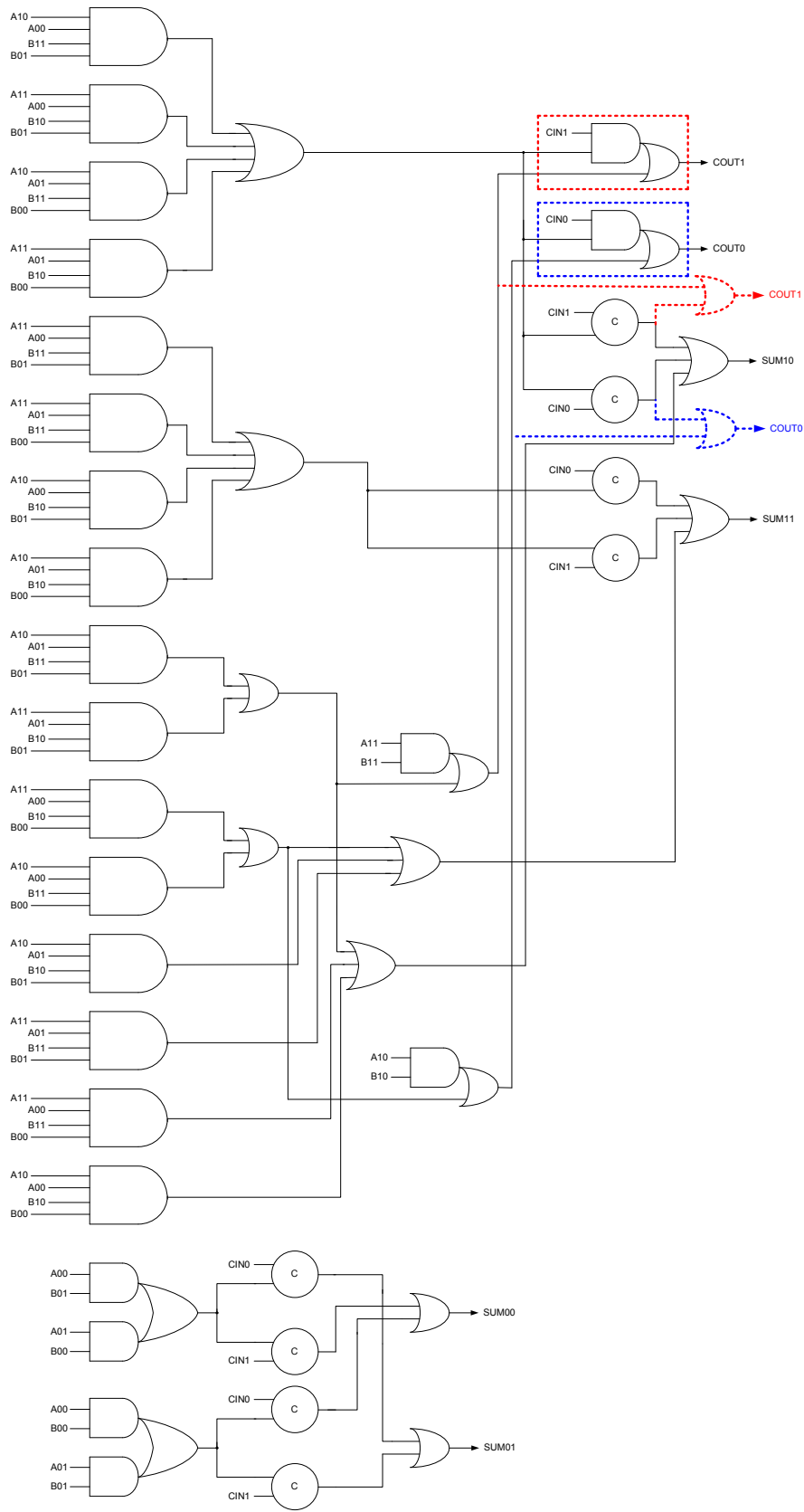

Fig 3 Proposed asynchronous early output DBFA(s) based on homogeneous delay-insensitive data encoding employing the 1-of-2 code

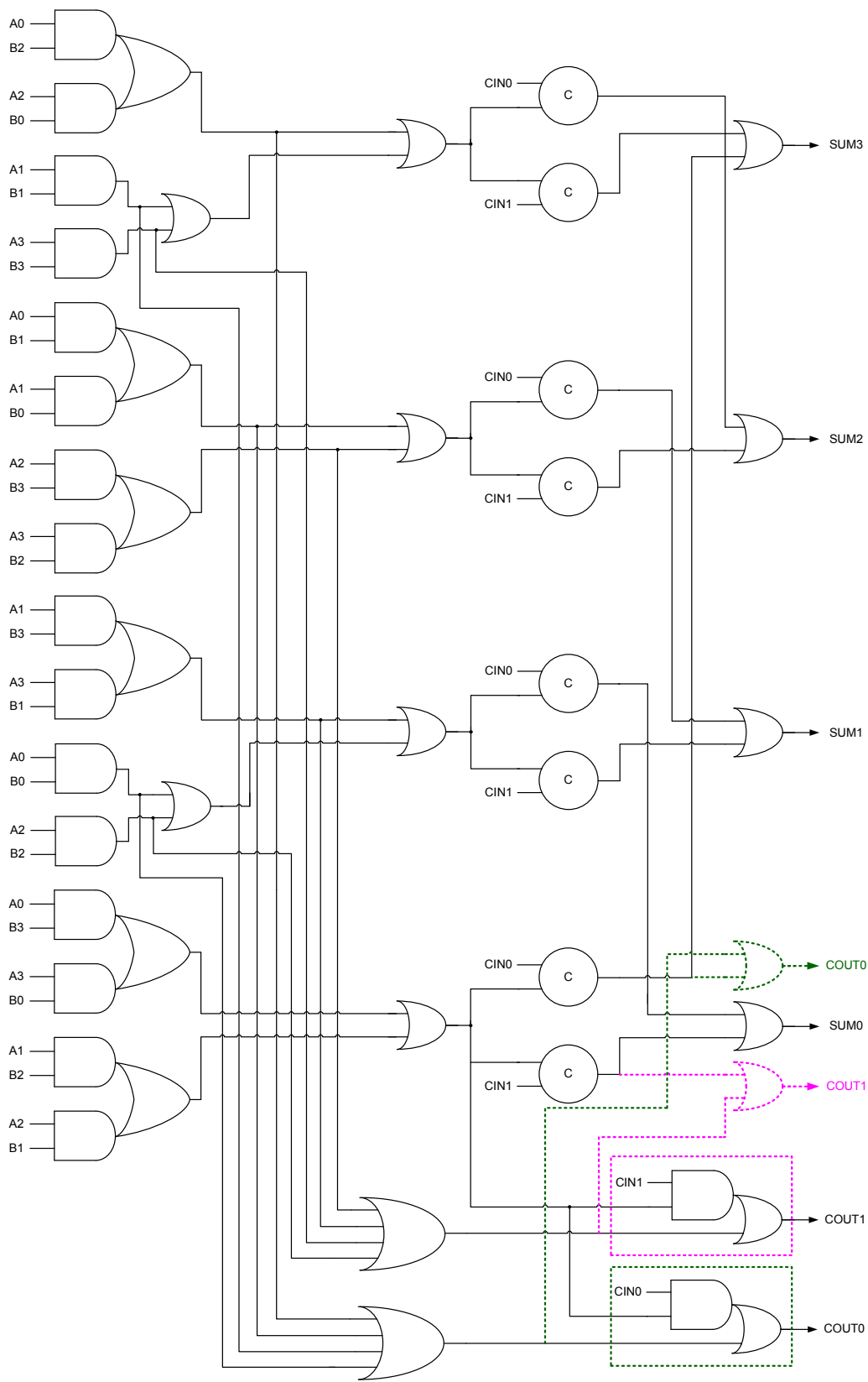

Fig 4 Proposed asynchronous early output DBFA(s) based on heterogeneous delay-insensitive data encoding employing 1-of-2 and 1-of-4 codes

## 4 Physical Realization and Results

Many 32-bit asynchronous RCAs were physically implemented by cascading homogeneously encoded and heterogeneously encoded asynchronous DBFAs corresponding to weak-indication and the proposed early output types separately. The generic architectures of the homogeneously encoded and heterogeneously encoded asynchronous RCAs are given in [37] and the reader is referred to the same for details. The RCAs were realized using the standard library cells of a 32/28nm CMOS process [36]. The 2-input C-element was alone manually realized using 12 transistors and it was made available to physically implement the various asynchronous RCAs. High fan-in C-element functionality wherever imminent was safely decomposed into a logic tree of 2-input C-elements using the quasi-delay-insensitive logic decomposition method presented in [38] which guarantees gate-orphan freedom.

An asynchronous system stage, as shown in Fig 1, consists of the asynchronous function block, the input registers and the completion detector. However for asynchronous function blocks realized on the basis of heterogeneous encoding, dual-rail to 1-of-4 encoders are introduced before the function block and 1-of-4 to dual-rail decoders are introduced after the function block as shown in [37]. In general, the input registers and the completion detector are identical and only the function blocks would in fact differ in their physical composition. Hence any differences between the simulation results of the various asynchronous RCAs can be attributed to the physical differences in their function block constituents. This paves the way for a straightforward comparison of the design metrics viz. latency, area and power of the different asynchronous RCAs post physical synthesis.

More than 1000 random input vectors were identically supplied to the various asynchronous RCAs at time intervals of 20ns through test benches to verify their functionalities and to capture their respective switching activities. The value change dump files generated were used for average power estimation using Synopsys tool. Since the EDA tool, by default, estimates the critical path timing, the worst-case forward latency was alone estimated for a typical case PVT specification viz. 1.05V and 25ºC of the standard cell library [36]. Default wire loads were automatically inserted while performing the simulations. A virtual clock was used to constrain the input and output ports of the asynchronous RCAs and it did not contribute to the actual power dissipation. Table 1 presents the simulation results obtained viz. critical path delay (also called forward latency), area occupancy, and average power dissipation for the different 32-bit asynchronous RCAs.

Table 1 Simulation results of various 32-bit asynchronous RCAs corresponding to weak-indication or early output incorporating diverse homogeneously or heterogeneously encoded DBFAs

| Asynchronous DBFA and Type | Latency (ns) | Area (µm$^2$) | Power (µW) |
|---|---|---|---|
| References [17,37]; No redundancy; Homogeneous encoding; Weak-indication | 4.12 | 2866.49 | 2200 |
| References [17,37]; Logic redundancy; Homogeneous encoding; Weak-indication | 2.84 | 2931.55 | 2202 |
| This work; No redundancy; Homogeneous encoding; Early output | 4.01 | 2472.06 | 2174 |
| This work; Logic redundancy; Homogeneous encoding; Early output | 2.21 | 2488.32 | 2173 |
| References [18,37]; No redundancy; Heterogeneous encoding; Weak-indication | 4.36 | 3301.58 | 2191 |
| References [18,37]; Logic redundancy; Heterogeneous encoding; Weak-indication | 3.03 | 3366.65 | 2192 |
| This work; No redundancy; Heterogeneous encoding; Early output | 4.22 | 2634.71 | 2182 |
| This work; Logic redundancy; Heterogeneous encoding; Early output | 2.38 | 2650.98 | 2182 |

In general, it can be inferred from Table 1 that the RCAs constituting homogeneously/heterogeneously encoded DBFAs with redundant logic facilitate good reductions in latency over their counterpart RCAs incorporating DBFAs with no redundant logic.

From Table 1, it is observed that compared to the weakly indicating 32-bit RCA incorporating the homogeneously encoded dual-bit full adder with redundant logic, the early output 32-bit RCA comprising the proposed dual-bit full adder with redundant logic that is based on homogeneous data encoding reports respective reductions in latency and area by 22.2% and 15.1% with no associated power penalty (in fact, a 1.3% power reduction results for the latter). Further, in comparison with the recently proposed early output 32-bit asynchronous carry select adder [39] that corresponds to the uniform input partition 8-8-8-8 and which is based on homogeneous dual-rail data encoding, the 32-bit asynchronous RCA incorporating the proposed early output dual-bit full adder with redundant logic reports 10.2% less latency, occupies 17.1% less area, and dissipates 5.2% less power.

On the other hand, the early output 32-bit RCA incorporating the proposed heterogeneously encoded dual-bit full adder with redundant logic reports corresponding decreases in latency and area than the weakly indicating 32-bit RCA that incorporates the heterogeneously encoded dual-bit full adder with redundant logic by 21.5% and 21.3% with nil power overhead (in fact, a 0.5% power reduction results for the former). Hence, overall, the early output 32-bit asynchronous RCA incorporating the proposed dual-bit full adder with redundant logic that is based on homogeneous data encoding is preferable.

## 5 Conclusion

This paper has presented new asynchronous early output DBFA designs based on homogeneous and heterogeneous delay-insensitive data encodings which when used to construct robust early output asynchronous RCAs lead to optimized design metrics compared to the weak-indication RCA counterparts constructed using weakly indicating asynchronous DBFAs. Overall, the simulation results show that the early output asynchronous RCAs constructed using homogeneously encoded DBFAs which have logic redundancy facilitate simultaneous optimizations in latency, area and power dissipation. Future work may consider evaluating the benefits of the proposed early output DBFAs incorporating redundant logic for asynchronous multi-operand additions [40].

*References:*
[1] S. Goel, A. Kumar, M.A. Bayoumi, "Design of robust, energy-efficient full adders for deep submicrometer design using hybrid-CMOS logic style," *IEEE Trans. on VLSI Systems*, vol. 14, no. 12, pp. 1309-1321, 2006.
[2] P. Balasubramanian, N.E. Mastorakis, "High speed gate level synchronous full adder designs," *WSEAS Trans. on Circuits and Systems*, vol. 8, no. 2, pp. 290-300, 2009.
[3] P. Balasubramanian, N.E. Mastorakis, "A delay improved gate level full adder design," *Proc. 3rd European Computing Conf.*, pp. 97-102, 2009.
[4] P. Balasubramanian, N.E. Mastorakis, "A low power gate level full adder module," *Proc. 3rd Intl. Conf. on Circuits, Systems and Signals*, Invited Paper, pp. 246-248, 2009.
[5] C.L. Seitz, "System Timing," in *Introduction to VLSI Systems*, C. Mead and L. Conway (Editors), pp. 218-262, Addison-Wesley, Reading, Massachusetts, USA, 1980.
[6] A.J. Martin, "Asynchronous datapaths and the design of an asynchronous adder," *Formal Methods in System Design*, vol. 1, no. 1, pp. 117-137, 1992.
[7] W.B. Toms, D.A. Edwards, "Efficient synthesis of speed independent combinational logic circuits," *Proc. 10th Asia and South Pacific Design Automation Conf.*, pp. 1022-1026, 2005.
[8] B. Folco, V. Bregier, L. Fesquet, M. Renaudin, "Technology mapping for area optimized quasi delay insensitive circuits," *Proc. Intl. Conf. on VLSI-SoC*, pp. 146-151, 2005.
[9] P. Balasubramanian, D.A. Edwards, "A delay efficient robust self-timed full adder," *Proc. 3rd IEEE Intl. Design and Test Workshop*, pp. 129-134, 2008.
[10] P. Balasubramanian, D.A. Edwards, "Self-timed full adder designs based on hybrid input encoding," *Proc. 12th IEEE Symp. on Design and Diagnostics of Electronic Circuits and Systems*, pp. 56-61, 2009.
[11] P. Balasubramanian, "A robust asynchronous early output full adder," *WSEAS Trans. on Circuits and Systems*, vol. 10, no. 7, pp. 221-230, 2011.
[12] P. Balasubramanian, "A latency optimized biased implementation style weak-indication self-timed full adder," *Facta Universitatis, Series: Electronics and Energetics*, vol. 28, no. 4, pp. 657-671, 2015.
[13] P. Balasubramanian, "An asynchronous early output full adder and a relative-timed ripple carry adder," *WSEAS Trans. on Circuits and Systems*, vol. 15, pp. 91-101, 2016.
[14] P. Balasubramanian, S. Yamashita, "Area/latency optimized early output asynchronous full adders and relative-timed ripple

carry adders," *SpringerPlus*, vol. 5:440, pages 26, 2016.

[15] P. Balasubramanian, K. Prasad, "Early output hybrid input encoded asynchronous full adder and relative-timed ripple carry adder," *Proc. 14th Intl. Conf. on Embedded Systems, Cyber-physical Systems, and Applications*, pp. 62-65, 2016.

[16] P. Balasubramanian, K. Prasad, N.E. Mastorakis, "A standard cell based synchronous dual-bit adder with embedded carry look-ahead," *WSEAS Trans. on Circuits and Systems*, vol. 9, no. 12, pp. 736-745, 2010.

[17] P. Balasubramanian, D.A. Edwards, "Dual-sum single-carry self-timed adder designs," *Proc. IEEE Computer Society Annual Symp. on VLSI*, pp. 121-126, 2009.

[18] P. Balasubramanian, D.A. Edwards, "Heterogeneously encoded dual-bit self-timed adder," *Proc. IEEE Ph.D. Research in Microelectronics and Electronics Conf.*, pp. 120-123, 2009.

[19] J. Sparsø, S. Furber, *Principles of Asynchronous Circuit Design: A Systems Perspective*, Kluwer Academic Publishers, Boston, MA, USA, 2001.

[20] T. Verhoeff, "Delay-insensitive codes – an overview," *Distributed Computing*, vol. 3, no. 1, pp. 1-8, 1988.

[21] C. Jeong, S.M. Nowick, "Optimization of robust asynchronous circuits by local input completeness relaxation," *Proc. Asia and South Pacific Design Automation Conf.*, pp. 622-627, 2007.

[22] P. Balasubramanian, K. Prasad, N.E. Mastorakis, "Robust asynchronous implementation of Boolean functions on the basis of duality," *Proc. 14th WSEAS Intl. Conf. on Circuits*, pp. 37-43, 2010.

[23] P. Balasubramanian, *Self-Timed Logic and the Design of Self-Timed Adders*, PhD thesis, The University of Manchester, 2010.

[24] A.J. Martin, "The limitation to delay-insensitivity in asynchronous circuits," *Proc. 6th MIT Conf. on Advanced Research in VLSI*, pp. 263-278, 1990.

[25] B. Bose, "On unordered codes," *IEEE Trans. on Computers*, vol. 40, no. 2, pp. 125-131, 1991.

[26] S.J. Piestrak, T. Nanya, "Towards totally self-checking delay-insensitive systems," *Proc. 25th Intl. Symp. on Fault-Tolerant Computing*, pp. 228-237, 1995.

[27] V.I. Varshavsky (Ed.), *Self-Timed Control of Concurrent Processes: The Design of Aperiodic Logical Circuits in Computers and Discrete Systems*, Chapter 4: Aperiodic Circuits, pp. 77-85, (Translated from the Russian by A.V. Yakovlev), Kluwer Academic Publishers, 1990.

[28] P. Balasubramanian, D.A. Edwards, "Efficient realization of strongly indicating function blocks," *Proc. IEEE Computer Society Annual Symp. on VLSI*, pp. 429-432, 2008.

[29] P. Balasubramanian, D.A. Edwards, "A new design technique for weakly indicating function blocks," *Proc. 11th IEEE Workshop on Design and Diagnostics of Electronic Circuits and Systems*, pp. 116-121, 2008.

[30] P. Balasubramanian, N.E. Mastorakis, "Global versus local weak-indication self-timed function blocks – a comparative analysis," *Proc. 10th Intl. Conf. on Circuits, Systems, Signal and Telecommunications*, pp. 86-97, 2016.

[31] C.F. Brej, J.D. Garside, "Early output logic using anti-tokens," *Proc. 12th Intl. Workshop on Logic and Synthesis*, pp. 302-309, 2003.

[32] P. Balasubramanian, "Comments on "Dual-rail asynchronous logic multi-level implementation"," *Integration, the VLSI Journal*, vol. 52, no. 1, pp. 34-40, 2016.

[33] P. Balasubramanian, R. Arisaka, H.R. Arabnia, "RB_DSOP: A rule based disjoint sum of products synthesis method," *Proc. 12th Intl. Conf. on Computer Design*, pp. 39-43, 2012.

[34] P. Balasubramanian, N.E. Mastorakis, "A set theory based method to derive network reliability expressions of complex system topologies," *Proc. Applied Computing Conf.*, pp. 108-114, 2010.

[35] P. Balasubramanian, D.A. Edwards, "Self-timed realization of combinational logic," *Proc. 19th Intl. Workshop on Logic and Synthesis*, pp. 55-62, 2010.

[36] Synopsys *SAED_EDK32/28_CORE* Databook, Revision 1.0.0, 2012.

[37] P. Balasubramanian, D.A. Edwards, W.B. Toms, "Redundant logic insertion and latency reduction in self-timed adders," *VLSI Design*, vol. 2012, Article ID 575389, pages 13, 2012.

[38] P. Balasubramanian, N.E. Mastorakis, "QDI decomposed DIMS method featuring homogeneous/heterogeneous data encoding," *Proc. Intl. Conf. on Computers, Digital Communications and Computing*, pp. 93-101, 2011.

[39] P. Balasubramanian, "Asynchronous carry select adders," *Engineering Science and Technology, an Intl. Journal*, 2017, DOI: http://dx/doi.org/10.1016/j.jestch.2017.02.003

[40] P. Balasubramanian, D.A. Edwards, W.B. Toms, "Self-timed multi-operand addition," *Intl. Jour. of Circuits, Systems and Signal Processing*, vol. 6, no. 1, pp. 1-11, 2012.